\documentclass{llncs}  

\usepackage{graphicx} 
\usepackage[T1]{fontenc}
\usepackage{amsmath}
\usepackage[english]{babel}
\usepackage{array,tabularx}
\usepackage{booktabs} 
\usepackage{makeidx}  
\usepackage{xparse}
\usepackage{moredefs}
\usepackage{lips}
\usepackage{epstopdf} 
\usepackage{color}
\usepackage{csquotes}
\usepackage{xcolor}
\usepackage{times}
\usepackage[hidelinks]{hyperref}
\usepackage[nolist,nohyperlinks]{acronym}
\usepackage{comment}
\usepackage{paralist}
\usepackage{cleveref}
\usepackage{wrapfig}
\usepackage{multirow}
\usepackage{microtype}
\usepackage{listings}
\usepackage[super]{nth}
\usepackage{natbib}

\usepackage{enumitem,amssymb}
\newlist{todolist}{itemize}{2}
\setlist[todolist]{label=$\square$}
\usepackage{pifont}
\usepackage{progressbar}

\usepackage{float}
\floatstyle{plaintop}
\restylefloat{table}

\usepackage[style=base,labelfont=bf,labelsep=period,tableposition=top,font=small]{caption}
\makeatletter
\renewcommand\@biblabel[1]{#1.}
\makeatother
\addto\captionsenglish{}

\usepackage{fancyhdr}
\fancypagestyle{WI_footer}{\fancyhf{}\fancyfoot[L]{\color{black}\scriptsize 20th International Conference on Wirtschaftsinformatik\\September 2025, Münster, Germany}}

\crefformat{footnote}{#2\footnotemark[#1]#3}
\expandafter\def\expandafter\UrlBreaks\expandafter{\UrlBreaks
  \do\a\do\b\do\c\do\d\do\e\do\f\do\g\do\h\do\i\do\j%
  \do\k\do\l\do\m\do\n\do\o\do\p\do\q\do\r\do\s\do\t%
  \do\u\do\v\do\w\do\x\do\y\do\z\do\A\do\B\do\C\do\D%
  \do\E\do\F\do\G\do\H\do\I\do\J\do\K\do\L\do\M\do\N%
  \do\O\do\P\do\Q\do\R\do\S\do\T\do\U\do\V\do\W\do\X%
  \do\Y\do\Z}

\newcolumntype{L}[1]{>{\raggedright\arraybackslash}p{#1}}   
\newcolumntype{C}[1]{>{\centering\arraybackslash}p{#1}}     
\newcolumntype{R}[1]{>{\raggedleft\arraybackslash}p{#1}}    

\begin{document}
\frontmatter          

\mainmatter              

\title{A Multi-Level Strategy for Deepfake Content Moderation under EU Regulation}

\subtitle{Research Paper} 


\author{
Max-Paul Förster\inst{1} \and
Luca Deck\inst{1,2,3} \and
Raimund Weidlich\inst{1} \and
Niklas Kühl\inst{1,2,3}}

\institute{University of Bayreuth, Germany\\
\email{luca.deck@uni-bayreuth.de} \and
Fraunhofer FIT, Bayreuth, Germany \and
FIM Research Center, Bayreuth, Germany
}

\maketitle
\setcounter{footnote}{0}

\begin{abstract}
The growing availability and use of deepfake technologies increases risks for democratic societies, e.g., for political communication on online platforms. The EU has responded with transparency obligations for providers and deployers of Artificial Intelligence (AI) systems and online platforms. This includes marking deepfakes during generation and labeling deepfakes when they are shared. However, the lack of industry and enforcement standards poses an ongoing challenge.
Through a multivocal literature review, we summarize methods for marking, detecting, and labeling deepfakes and assess their effectiveness under EU regulation. Our results indicate that individual methods fail to meet regulatory and practical requirements. Therefore, we propose a multi-level strategy combining the strengths of existing methods. To account for the masses of content on online platforms, our multi-level strategy provides scalability and practicality via a simple scoring mechanism.
At the same time, it is agnostic to types of deepfake technology and allows for context-specific risk weighting.

\end{abstract}

\thispagestyle{WI_footer}






\section{Introduction}
\label{sec:introduction}
Due to rapid progress in the development and availability of artificial intelligence (AI) technologies, deepfakes have found their way into modern society. 
Deepfakes are AI-generated media content pretending to resemble real persons, objects or events~(Art. 3(60) AI Act) and cover a wide variety of applications ranging from artistic use to disinformation. 
The US elections in 2024, but also current geopolitical conflicts, show that deepfakes are used to manipulate and divide society \citep{Byman2023InternationalConflicts}.
In Germany, a deepfake video with former chancellor Olaf Scholz announcing a ban of the Alternative für Deutschland (AfD) has caused huge media attention~\citep{ScholzDeepfakeOriginal.2023, ScholzDeepfakeNetzpolitik.2023}.
In its global risk report 2025, the Word Economic Forum (WEF) categorized risks of misinformation and disinformation in combination with deepfake technologies as the most relevant risks for the coming years~\citep{WEF2025RiskReport}.
To mitigate these risks, the European Union (EU) has passed regulations requiring providers and deployers of AI systems as well as online platforms to make deepfakes transparent. 
However, enforcing these regulations remains difficult as practitioners and enforcement agencies face significant challenges ~\citep{Demková.Giovanni}.
First, deepfake technology is rapidly evolving, making reliable detection technically difficult and costly---particularly with huge amounts of content.
Second, handling deepfakes is subject to intricate tradeoffs between economic and societal risks, such as efficiency and freedom of speech.
At the same time, practical guidance is still scarce.
In this paper, we combine expertise from law and IS to foster efficient and effective content moderation of deepfakes under EU regulation. Having an interdisciplinary perspective is essential. IS research provides insights into the technical and economic feasibility of methods to mitigate risks of deepfakes, while legal scholarship addresses accountability and governance, ensuring a proportionate approach.
This unique setup allows for a twofold contribution: \textit{First}, we analyze to which extent existing methods are suitable to mitigate societal risks of deepfakes specified under EU regulation. \textit{Second}, we propose a novel multi-level strategy for detecting deepfakes that is both practical and flexible enough to be employed for content moderation on online platforms.
The remainder of this work is organized as follows. We start by summarizing societal risks connected to deepfakes and introducing relevant EU regulation on deepfakes in \Cref{sec:background}.
Based on a multivocal literature review described in \Cref{sec:method}, \Cref{sec:results} summarizes existing methods for implementing the requirements of EU regulation.
\Cref{sec:strategy} synthesizes the strengths and weaknesses of the identified methods into a multi-level strategy that can be adopted by online platforms and enforcement agencies. 
We also discuss context-specific tradeoffs and challenges in \Cref{sec:discussion} and conclude with limitations and prospects of our multi-level strategy in \Cref{sec:conclusion}.

\section{Background}
\label{sec:background}
Societal risks connected to deepfakes have prompted the EU to pass several regulatory acts to face and mitigate dangers of deepfakes, including the Artificial Intelligence Act (AI Act) and the Digital Services Act (DSA). Here, we summarize societal risks of deepfakes and introduce relevant EU regulation to mitigate these risks.

\subsection{Societal Risks of Deepfakes}
Actors who instrumentalize deepfakes to influence the individual and public opinion pose a substantial threat to society~\citep{Karaboga2024}.  
The deceptive nature of deepfakes makes it particularly difficult for individuals---especially with low digital literacy---and the public to recognize this content, resulting in unprecedented and hardly foreseeable risks for society. 
This is particularly evident in the context of digital structural changes in public communication and information strategy \citep{Karaboga2024}.   
Nowadays, public debates are not only shaped by trusted news agencies who publish curated news content, but also by private individuals with vast reach via online platforms \citep{Jarren2021,Karaboga_2023}.   
The gatekeeper function of the ``traditional'' media no longer exists in its historic form \citep{Flamme2024,Eisenegger2019}.
Individuals can now be reached directly without any controlling body \citep{Habermas_2022,Karaboga2024}.   
Although this facilitates deliberation in the opinion-forming process, it also leads to uncontrolled distribution of content and a fading of necessary rules for public debate \citep{Habermas_2022}. This provides the potential for both foreign state and non-state actors or political parties to strategically manipulate the public opinion \citep{Kleemann2023,Karaboga2024}. 
These potential threats to society and democracy are a main driver for the EU regulation (Recital 1 AI Act).

\subsection{EU Regulation on Deepfakes}
\vspace{-10pt}
\begin{figure}
    \centering
    \includegraphics[width= 1\linewidth]{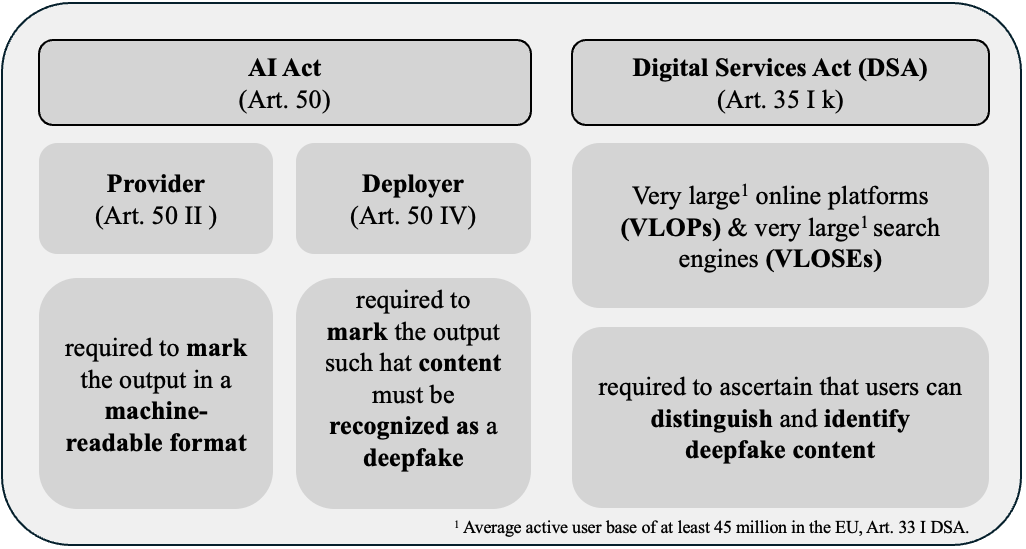}
    \caption{Overview of EU regulation to mitigate dangers of deepfakes.}
    \label{fig:enter-label}
\end{figure}
The AI Act aims to regulate AI systems in the EU and their providers and deployers through a risk-based approach (\citealp{VonWelser2024}; Art. 2 AI Act).
Starting from August 2026, it obliges providers to disclose that the content has been generated or was manipulated by AI-systems (\citealp{VonWelser2024}; Art. 50(2), 50(4) AI Act). 
Art. 50 of the AI Act divides the transparency obligation for deepfakes between the use cases of providers (Art. 50(2) AI Act) and deployers (Art. 50(4) AI Act). 
A provider of AI-systems who generates deepfakes is required to \textit{mark}\footnote{To avoid ambiguity, we refer to information inherent to the content as ``mark'', whereas we refer to information added by online platforms as ``label''.} the output in a machine-readable format (Art. 50(2) AI Act).
Additionally, deployers of AI systems who generate deepfakes are required to \textit{mark} the output such that content must be recognized as a deepfake.
One of the reasons for this requirement is to facilitate compliance with the transparency obligations under the DSA (\citealp{Merkle2024}; AI Act Recitals 136).
The DSA focuses on the regulation of online intermediary services, such as social media platforms and search engines, that offer their services to users in the EU (Art. 1 DSA).
In the context of deepfake regulation, the most relevant transparency obligations only apply to the highest level (Art. 35(1)(k) DSA), i.e., very large online platforms (VLOPs) and very large search engines (VLOSEs). 
VLOPs and VLOSEs need to \textit{label} content which contains deepfakes (Art. 35(1)(k) DSA).
\subsection{Research Gap}
In light of recent technological and legal advances, growing attention is being directed towards interpretations of EU regulation for ``ethical AI''~\citep{Siau.2020}, sparking debates on specific requirements like transparency \citep{Holst2024, Hacker.Varietiesof2022} and fairness~\citep{Wachter.Whyfairness2021, Deck.Implicationsof2024}.
Prior work has also started to address the practical implications of deepfake regulation.
\citet{Karaboga2024} provide an extensive impact assessment of deepfakes and derive practical recommendations to a range of stakeholders, including online platforms, regulators, and society. This work particularly highlights implications for political communication and presents an essential basis for our work.
However, it focuses primarily on the Swiss judicial area (\citealp{Karaboga2024}).
On the other hand, a range of studies (e.g., \citealp{MolaviVasse2024, MolaviUdoh2024}) respond to EU regulation and propose specific techniques such as watermarking to meet regulatory requirements.
However, the majority of the state-of-the-art marking and detection methods rely on a single point of assessment, with none of them presenting a one-size-fits-all solution (see \Cref{subsec:methods}).
This calls for a combination of methods into more robust strategies. For example, \citet{Gandhi2024} integrate both graphical and auditory elements into a multimodal detection approach. 
Moreover, multimodal approaches are also already used in traditional misinformation detection \citep{Zeng2024}.

\section{Method} \label{sec:method}
To fill in the gap outlined in \Cref{sec:background}, we surveyed existing methods for marking and detecting deepfakes and analyzed whether they are suitable to mitigate risks of deepfakes under EU regulation.
To this end, we conducted a multivocal literature review~\citep{Garousi2019}. This method accounts for the rapid recent developments surrounding technical and legal aspects of deepfakes by including grey literature~\citep{Schöpfel.Farace2010} such as preprints and official guidelines.
Our literature review strictly followed the central research question: ``\textit{How can online platforms effectively and efficiently handle deepfakes to mitigate the societal risks specified in EU regulation?}''
To identify relevant literature, we defined a search string consisting of four key elements: regulation, detection, deepfake, and political communication. 
For each element, we accounted for several variations, synonyms, and related terms.
This led to the following search string\footnote{For the German Beck and Juris databases, we used a translated and slightly shortened search string.}:

\vspace{-10pt}

\begin{center}
\footnotesize
\fbox{%
    \parbox{1\columnwidth}{
    \centering
        (Act OR Legislation OR Law OR Regulation OR Compliance OR Standards OR Detection)
    \textbf{AND} (deepfake OR deep fake OR synthetic media OR manipulated content OR disinformation OR AI OR Artificial Intelligence OR Machine Learning OR political OR political communication OR misinformation OR Election campaign OR Electoral Integrity OR Democracy)
    }}
\end{center}

\begin{figure}
    \centering
    \includegraphics[width=1\linewidth]{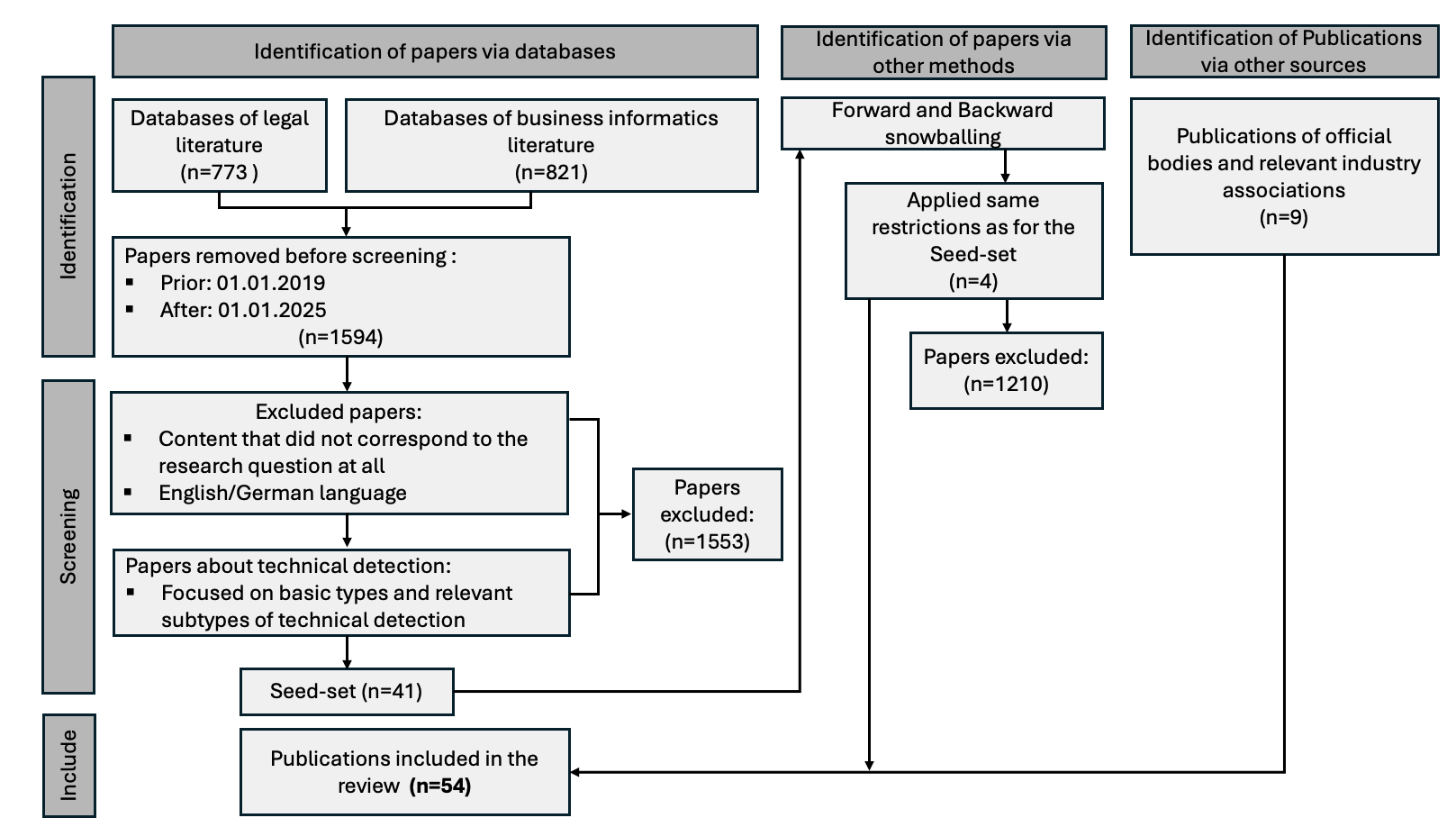}
    \caption{PRISMA flowchart describing the source selection procedure.}
    \label{fig:litrev}
\end{figure}

To combine both the relevant literature on EU regulation as well as computer science and IS, we applied the search string to the databases Scopus, arXiv, Beck, and Juris.
The application of the search strings across all databases yielded a total of 1.594 results (dataset-1) published between 01.01.2019 to 01.01.2025. 
To narrow down the relevant papers, we employed a filtering procedure, which is illustrated in \Cref{fig:litrev} following the PRISMA standard for literature reviews~\citep{Page.2021}. 
We first excluded papers that were not available in German or English and papers whose titles and abstracts obviously did not correspond to the research question. 
Afterwards, we switched to more in-depth screening of abstracts and full texts to select the papers with explicit contributions to our research question. 
This process reduced a total of 41 papers, which served as a seed set for the further process.

To identify relevant papers beyond the literature included in the databases and the search strings, we relied on proven guidelines for iterative backward and forward snowballing~\citep{Wohlin.2022}.
Using the literature management tool \textit{Citationchaser} \citep{Haddaway.2022} for the seed set and the same filter criteria as before, we obtained four additional papers.
We further augmented the academic literature with publications from official bodies of the EU and the German federal authorities, as well as literature from relevant industry associations, yielding nine additional papers. 
Following the guidelines of~\citet{Garousi2019}, we evaluated the gray literature based on authorship, methodology, institutional background, objectivity, and outlet type.

\begin{table}[h]
    \footnotesize
    \centering
    \begin{tabular}{p{1.6cm} p{10.4cm}}
        \toprule
        \textbf{Category} & \textbf{References} \\
        \midrule
        \textbf{Regulatory Guidance} & 
        \cite{VonWelser2024, KlosTaylan2024, Merkle2024, RaueHeesen2024, EUCommission2024ListDSA, Scientificservice2024, EUCommission2024GuidelinesDSA, Hennemann2024, Bitkom2024, EU-Parliament2023AITopics, Karaboga_2023, GersdorfPaal2024, Wybitul2024, BuchheimSchrenk2024, TaegerPohl2024} \\
        \midrule
        \textbf{Marking Methods} &
        \cite{MolaviVasse2024, MolaviUdoh2024, BellaajOuni2019, NadimpalliRattani2024, Becker2024, Boenisch_2021, EUCommission2024GuidelinesDSA, Dobber2023}\\
        \midrule
        \textbf{Technical Detection} & 
         \cite{MirskyLee2021, Agrawal2019, Kaur2024, Mittal2020, Marra2018, QiLuo2022, Le2023, Baier2023, Gambín2024, EUCommission2024GuidelinesDSA, ZhouZafarani2020} \\
        \midrule
        \textbf{Trusted Detection } & 
         \cite{Jaroucheh2020, Groh2021, EUCommission2024GuidelinesDSA, Becker2024, JarrahiSafari2021, Sängerlaub2020, Boenisch_2021, Allen2021}\\
        \midrule
        \textbf{Risks and Limitations} &
         \cite{EUCommission2024GuidelinesDSA, EU-Parliament2023AITopics, GersdorfPaal2024, ScholzDeepfakeNetzpolitik.2023, Dobber2023, Karaboga2024, Wybitul2024, WEF2025RiskReport, BuchheimSchrenk2024, Gambín2024, US-DepartmentofState2022PressBriefing, Brömmelhörster2024, Byman2023InternationalConflicts, ChesneyCitron2019, Jarren2021, Eisenegger2019, VanHuijtee2021, MetaPlatforms2024Statista, EESC2024, Habermas_2022, Hitzler2025, Kay2020, Kleemann2023, Simon2023, ZhouZafarani2020} \\
         
        \bottomrule
    \end{tabular}
    \caption{Overview of the results of the multivocal literature review.}
    \label{tab:literature_overview}
\end{table}

\section{Results} \label{sec:results}
The final set of 54 papers is organized into three categories, as depicted in \Cref{tab:literature_overview}.
\textit{Regulatory Guidance} comprises papers that explicitly engage with regulatory requirements, while \textit{Methods} include methods to handle deepfakes, divided into marking, detection, and trusted methods. \textit{Limitations and Risks} includes papers that address technical and practical limitations with regard to EU regulation, and the risks connected to these limitations. The following section summarizes regulatory requirements under the AI Act and the DSA and analyzes the suitability of existing methods for effective and efficient fulfillment of these requirements.

\subsection{Regulatory Guidance}
The recitals of the AI Act propose a range of technical methods to mark deepfakes, such as metadata, frequency component, cryptographic, and statistical watermarking (AI Act Recital 133).
In selecting a technical method, providers should consider the limitations of each method (Art. 50(2) AI Act). 
Furthermore, the deployer must mark the output in a way that discloses the artificially generated origin (Art. 50(4) AI Act).
This can be accomplished, e.g., using banners, overlays, popups, or auditory cues \citep{Bitkom2024}.  

The implementation of the labeling obligations under Art. 35 (1)(k) of the DSA presents a more significant challenge for the obligated parties than the implementation of the transparency obligations under Art. 50 of the AI Act. 
This is because VLOPs cannot trace the origin of content from third parties (users) with sufficient certainty. For example, users outside of the EU have access to technology that is not subject to the transparency obligations under the AI Act, and even within the EU, there is no guarantee that users comply with these transparency obligations.
Before labeling deepfakes, VLOPs must therefore implement detection frameworks that can distinguish real content from deepfakes. Consequently, a robust detection framework for deepfakes is a fundamental prerequisite for the successful implementation of the transparency obligations under Art. 35(1)(k) of the DSA.

\subsection{Marking and Detection Methods} \label{subsec:methods}
\paragraph{Marking Methods.}
As part of the implementation of the AI Act, the marking methods mentioned above are of particular relevance. Metadata watermarking embeds the mark directly in the metadata \citep{MolaviVasse2024}.  
This method has the advantage of not changing the content itself (e.g., in terms of quality;~\citealp{MolaviVasse2024}).  
However, this method is vulnerable to manipulation, in particular through the automatic removal of the metadata \citep{MolaviUdoh2024}.  
A complex but more robust method is frequency component watermarking, which includes a watermark in the frequency domain of a digital signal~\citep{BellaajOuni2019, MolaviUdoh2024}.  
Similarly, cryptographic watermarking is robust to manipulation by integrating a signature into the content itself \citep{MolaviVasse2024}.  
However, it cannot be applied to text content \citep{MolaviUdoh2024}.  
Statistical watermarking embeds a statistical pattern into the content, e.g., by changing pixel values \citep{MolaviUdoh2024}. This watermark is not detectable by the human eye and constitutes a robust method for labeling content \citep{MolaviVasse2024}. 
Notably, none of the watermarking methods can withstand all forms of manipulation attacks \citep{MolaviUdoh2024}.  
Moreover, the methods need to be constantly adapted to the evolving landscape of deepfake technologies and manipulation attacks to ensure long-term robustness. 

\paragraph{Technical Detection Methods.}
As previously outlined, deepfake detection methods are practically necessary to fulfill the obligations under Art. 35(1)(k) of the DSA. The large number of deepfake detection frameworks can be divided into different fundamental types with homogeneous strengths and weaknesses. 
One example of \textit{artefact-based methods} focuses on behavior. 
It uses monitoring of facial expressions and gestures to detect anomalies \citep{MirskyLee2021}.  
Although this method requires large amounts of training data, it has the potential to be an effective detection method for popular individuals such as politicians \citep{Agrawal2019}.  
However, the applicability of such approaches is limited \citep{Kaur2024}. First, they require the presence of specific artefacts \citep{Mittal2020}. Second, the restriction to a specific artefact or group of artefacts makes them easy to anticipate for detection bypasses \citep{Kaur2024,Le2023}. 

Unlike artefact-specific methods, \textit{undirected methods} do not rely on one pre-defined set of artefacts. 
Instead, the algorithm autonomously selects distinctive criteria, either via classification or anomaly detection~\citep{MirskyLee2021}.  
Classification-based methods rely on supervised deep learning models and labeled images to differentiate between real content and deepfakes.
Although they can perform significantly better than artefact-specific methods \citep{Marra2018}, they also show limitations. 
First, the training of such models requires a large and representative amount of labeled deepfakes, which is difficult to obtain \citep{Kaur2024,QiLuo2022}. 
This makes them prone to concept drift~\citep{Baier2023,Kaur2024}, i.e., loss of performance due to changes in the environment (e.g., new deepfake technologies).  
Further, deep learning models are blackboxes, which makes it difficult to retrace and challenge the reasoning for classification decisions \citep{Kaur2024}. 
Another undirected approach is anomaly detection, which requires real images to detect differences from deepfakes \citep{MirskyLee2021}.  
However, due to the increasing amount of artificially generated or manipulated digital content, training data (i.e., real images) must be carefully curated to avoid mislabeled data \citep{MolaviVasse2024}. 

\paragraph{Trusted Methods.}
In contrast to fully technical detection methods, trusted methods use collective intelligence or human knowledge to detect deepfakes.  
\citet{Jaroucheh2020} differentiates between ``expert-based fact checking'' and ``crowd-sourced fact checking''.
Expert-based fact checking focuses not only on superficial patterns, but can also account for context~\citep{Jaroucheh2020}.  
For example, deepfakes might be recognized due to untrustworthy statements made in the deepfake itself.
However, independent and accepted experts are scarce \citep{Jaroucheh2020} and not sufficiently efficient for large amounts of content in a short period of time  \citep{Sängerlaub2020}, limiting its applicability for VLOPs
%
Crowd-sourced fact checking relies on collective intelligence of many individuals rather than expert individuals\citep{Jaroucheh2020}.   
For example, collective signing has a selected group of verifiers independently reviewing content \citep{Jaroucheh2020}.  
The advantages of these methods are scalability and diversity of expertise and knowledge~\citep{Jaroucheh2020}. 
Initial evidence suggests that representative crowds can classify disinformation just as accurately as experts~\citep{Allen2021}.  
However, it requires sufficient incentives for the verifiers \citep{Jaroucheh2020} and poses the risk of external manipulation \citep{Allen2021}.
Therefore, VLOPs would face the challenge to establish scalable systems with sufficient availability and performance in real-time \citep{Jaroucheh2020}.

\section{A Multi-Level Strategy for Deepfake Detection and Labeling} \label{sec:strategy}
With an understanding of the range of available methods and their individual strengths and weaknesses at hand, we combine the strengths into a multi-level strategy that can be adopted by online platforms and enforcement agencies to label potential deepfakes in a proportionate manner, to overcome the current systemic challenges of the EU. 

\subsection{Multi-Level Detection}
\begin{figure}
    \centering
    \includegraphics[width=1\linewidth]{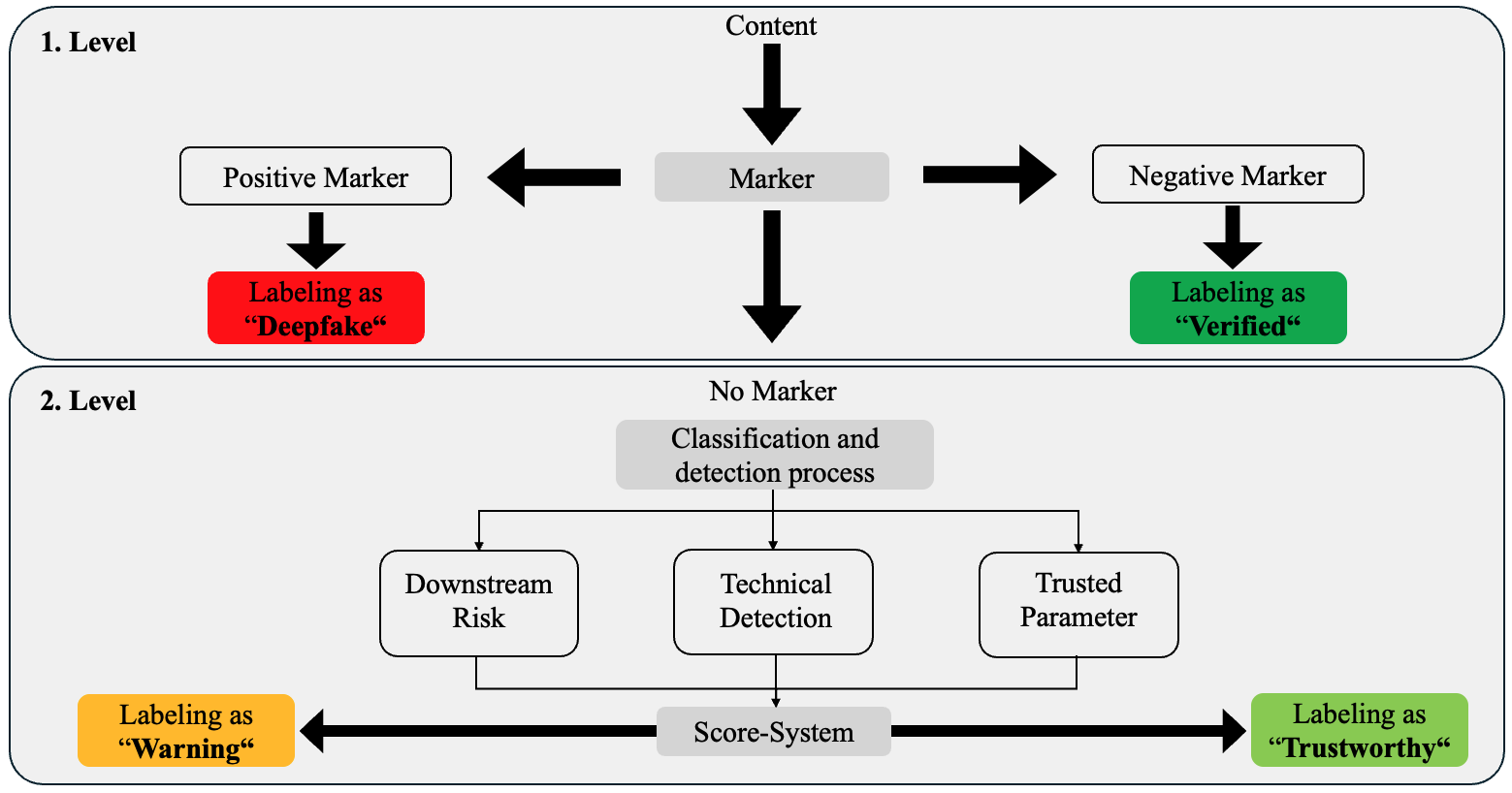}
    \caption{Multi-level strategy using markers and classification mechanisms to label potential deepfake content.}
    \label{fig:multi-level strategy}
\end{figure}

As online platforms cannot reliably expect deepfakes to be marked, two levels are required.
As depicted in \Cref{fig:multi-level strategy}, the first level pre-categorizes content based on markers that may be embedded under the AI Act.
The second level employs a more complex procedure to account for content without markers for efficient handling of difficult cases.

The first level checks markers such as watermarks (see \Cref{subsec:methods}). 
Positive markers mark the content as a deepfake and will be increasingly common due to the transparency obligations in the AI Act. 
As a counterpart, negative markers can authenticate content that has not been modified or created by artificial intelligence \citep{NadimpalliRattani2024, Becker2024}.  
Although negative markers are currently not widespread, it can be anticipated that they will play a more significant role in the future, e.g., due to new regulations and industry standards. 
Crucially, markers---in particular negative markers---require a verified certification chain to prevent potential tampering \citep{Boenisch_2021}.

In the absence of markers, e.g., when content originates from outside the EU or is intentionally deceptive~\citep{Kleemann2023,ChesneyCitron2019}, deepfakes need to be detected.
To combine the individual strengths of existing methods, we propose a multimodal approach that combines methods from both detection categories identified in \Cref{subsec:methods}, i.e., technical detection and trusted detection.
While technical detection methods are highly scalable and sensitive to minor details, trusted detection methods can account for the context of the content.
Also, the detection process accounts for potential downstream risks, which have to be classified based on the content and its context, such as origin and potential reach. This is particularly relevant when outcomes of individual detection methods differ or a clear result cannot be achieved quickly enough.
The systemic risks described in Art. 34(1)(c) of the DSA could be used to classify downstream risk, in particular threats to our democratic society, such as political communication and catastrophic events. 
Similar to deepfake detection, the downstream risk of specific content can be classified using technical approaches or human expertise.
The exact selection and configuration of methods, as well as the classification of downstream risk, depend on the contextual requirements. Still, we shortly sketch how our multi-level strategy might be instantiated for a VLOP.
As technical methods, undirected methods are broadly applicable and highly scalable to detect a majority of deepfakes with high certainty. 
Trusted methods become particularly relevant when technical methods fail or the content embodies high downstream risk~\citep{Gambín2024}.
Similar to Google's reCAPTCHA \citep{recaptcha}, VLOPs could create collective signing systems for paid, voluntary, or mandatory content moderation. Additionally, they could install experts for intricate cases where average people struggle \citep{Groh2021}.
With regard to downstream risk, content on VLOPs could be classified into harmless categories, such as animals, and potentially harmful categories, such as political communication.
In addition, selected content providers, such as media outlets following certain journalistic standards, could be verified \citep{EUCommission2024GuidelinesDSA, Becker2024, JarrahiSafari2021}, to add more context during the detection procedure.  
\subsection{Scoring and Labeling}
\begin{figure}
    \centering
    \includegraphics[width=1\linewidth]{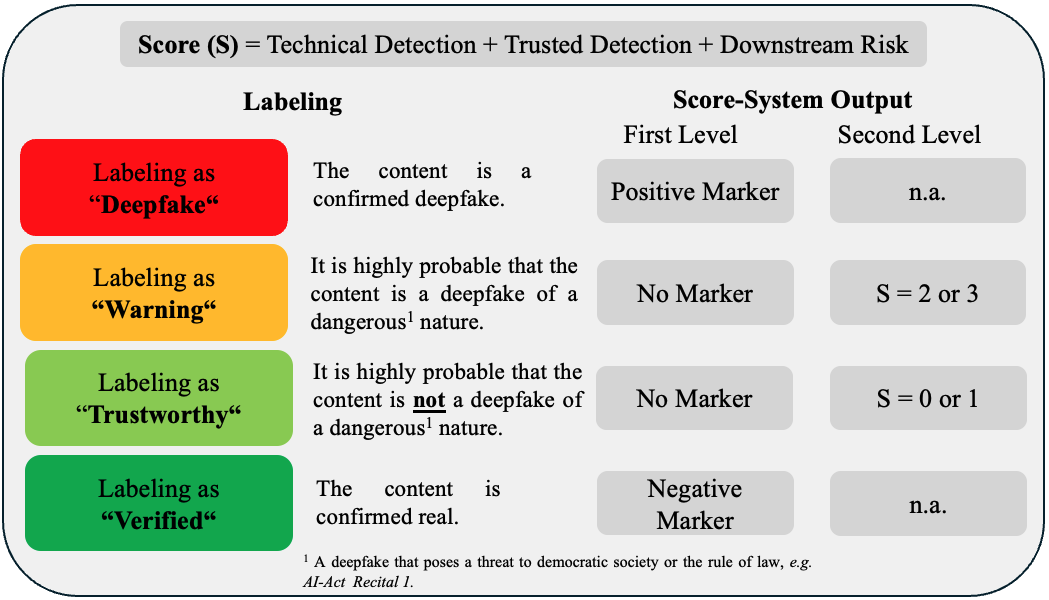}
    \caption{Exemplary scoring and labeling system based on our multi-level strategy.}
    \label{fig:labeling}
\end{figure}
As the ultimate goal is not detection but labeling of deepfakes, our multi-level strategy also guides decisions on how to assign labels.
In order to make a final decision on the classification of the content, a score system should combine three aspects: technical detection, trusted detection, and downstream risk. 
The exemplary score system in \Cref{fig:multi-level strategy} assigns binary values (0 if untrustworthy and 1 if trustworthy) and weights all three values equally.
These values are combined into an aggregate score that enables justifiable decisions about the classification and labeling of content.
For more fine-granular classification, the scoring can be refined, e.g., by modifying the weights based on criteria such as technical reliability or expertise.
The multimodal detection approach results in four potential labeling levels as outlined in \Cref{fig:labeling}. 
%
%
Positive markers are directly labeled as ``Deepfakes''. 
Contrarily, negative markers warrant a ``Verified''label \citep{EUCommission2024GuidelinesDSA},  \citep{Dobber2023}.  
The rest of the labels depend on the result of the score system from \Cref{fig:multi-level strategy}. 
Such transparent scoring and labeling could lead to more trust in the content moderation process and the content itself \citep{Dobber2023}. 

\section{Discussion} \label{sec:discussion}
By combining the merits of multiple approaches, our multi-level strategy could be more robust against the manifold types and risks of deepfakes. Importantly, however, the strategy must be instantiated based on context-specific requirements regarding economic practicality and legal compliance.
Here, we shed light on three key challenges and embed them in a broader discourse.
\subsection{Misclassification}
A high amount of mislabeled content is expected to lead to a loss of trust in the classification system \citep{Karaboga2024}. Therefore, it is paramount to minimize misclassifications by means of technical state-of-the-art and a combination of methods.
Still, no deepfake detection system will ever guarantee perfect accuracy, leading to intricate tradeoffs between types of misclassifications. In this context, actual deepfakes classified as verified or trustworthy are regarded as \textit{false negatives}, while real content classified as deepfake or untrustworthy presents a \textit{false positive}. 
Notably, the scoring system depicted in \Cref{fig:labeling} allows online platforms to trade-off false positives with false negatives by setting a more or less sensitive threshold and weighting of the three values.
It does, however, not specify how to set this threshold. This decision depends on context-specific priorities and needs to consider the respective costs of false positives and false negatives.
False negatives (deepfakes labeled as verified content) entail the very risks to our democratic society that the AI Act aims to mitigate, i.e., the spread of misinformation or even intentional disinformation campaigns. At the same time, singular false positives without high downstream risk or low reach are less costly, making the potential costs difficult to estimate.
On the other hand, costs of false positives (real content labeled as deepfake) entail unwarranted loss of trust in the specific content and reputational damage for the source of the content \citep{Dobber2023}. False positives may also severely limit freedom of speech or artistic freedom, posing another risk to our democratic society  \citep{Kozyreva}.
Future research must examine how to determine these tradeoffs. Costs of misclassifications might be further decreased by establishing fallback strategies that, e.g., allow for labeling or banning deepfakes even after they have been published on an online platform. Rare cases of critical downstream risk and high uncertainty might also warrant expert assessments, which may include human-AI collaboration combining technical methods with human expertise.


\subsection{Specification of Labeling Obligations}
A major question arises regarding the scope and limits of the labeling obligation. 
Art. 35(1)(k) of the DSA and Art. 50(4) in connection with Recital 133 of the AI Act are both framed very vaguely and constantly refer to the possible state-of-the-art.  
From the perspective of online platforms, however, clarification is needed on whether verification of real content falls under the obligations of the regulations. 
Art. 50(4) of the AI Act only provides labeling obligations for providers and deployers of AI systems that generate deepfakes. 
On the other hand, Art. 35(1)(k) of the DSA demands “\textit{reasonable, proportionate and effective mitigation measures, tailored to the specific systemic risks}”.
In addition, the EU Commission's guidelines already point toward cases such as scientific research where sources should be actively labeled as trustworthy \citep{EUCommission2024GuidelinesDSA, Dobber2023}.
One could therefore argue that in some cases, only multi-level labeling systems constitute ``\textit{effective mitigation measures}” under Art. 35(1)(k) of the DSA, particularly if one is aware of the regulation goals, which also aim to ensure a “\textit{safe, predictable and trusted online environment}” (Art. 1(1) DSA). 
As discussed before, trust in the content on VLOPs can only be effectively established through the classification and labeling of verified and trustworthy content in combination with a warning against deepfakes and untrustworthy content. 
Against this backdrop, even if explicit obligations are currently debatable, we expect multi-level labeling systems to become more relevant for types of content posing high potential downstream risk, such as political communication.

\subsection{Enforcement of the Labeling Obligations }
Lastly, for the transparency obligations under EU regulation to become effective, they will require proactive enforcement.
In the context of VLOPs, the EU Commission is specifically responsible for the enforcement (Art. 56(2), 56(3), 49(4) DSA), which is to be carried out by an independent auditing organization (Art. 37(1)(a), 37(3) DSA).
Random sample tests may be a feasible option to evaluate the performance of content moderation systems employed by VLOPs.
Nonetheless, effective enforcement requires auditing authorities to have an understanding of practical solutions themselves.
Ultimately, we encourage them to collaborate on context-specific applications of our multi-level strategy and to test different configurations of detection methods and score systems. 

\section{Conclusion} \label{sec:conclusion}
As the origin of digital content can not be reliably controlled, and deepfake detection is a moving target, online platforms and enforcement agencies alike are struggling to put the legal requirements of DSA and AI Act into practice.
Combining the strengths of existing detection methods, we propose a multi-level strategy for online platforms to efficiently label content and effectively mitigate societal risks under EU regulation.
While our proposal could be more holistic and robust than existing approaches, future work must evaluate whether our proposal holds in real application settings. The multi-level strategy has not been evaluated by external experts or tested and is merely a blueprint that does not specify concrete methods, risk weightings or implications for action.
We also leave many related questions for future work such as proportionality with regard to freedom of expression and other fundamental rights from the EU charter.
We perceive our multi-level strategy to be a building block in the larger endeavor to make democratic society resilient to the risks of deepfakes. 
A fully holistic solution will---among others---require a better understanding of the societal risks of deepfakes on online platforms \citep{al2023impact}, substantial efforts to improve media literacy of users \citep{EESC2024}, and more interdisciplinary collaboration for practical solutions. 

\section{Acknowledgements}
Work on this paper has been supported by the project „For the Greater Good? Deepfakes in Criminal Prosecution (FoGG)“ funded by the Bavarian Institute for Digital Transformation (bidt), an institute of the Bavarian Academy of Sciences and Humanities, under project code KON-024-008.


\bibliographystyle{agsm}
\bibliography{literature}

\end{document}